\documentclass[aps,prl,twocolumn,showpacs,amsmath,amssymb,amsfonts, long,preprintnumbers,superscriptaddress,nofootinbib]{revtex4}
\usepackage{graphicx}

\renewcommand{\section}[1]{{\textsl{#1}.}}

\begin{document}

\title{Revisiting the bound on axion-photon coupling from Globular Clusters}
\author{Adrian Ayala}
\affiliation{Universidad de Granada, 18071 Granada, Spain}
\author{Inma Dom\'inguez}
\affiliation{Universidad de Granada, 18071 Granada, Spain}
\author{Maurizio Giannotti}
\affiliation{Physical Sciences, Barry University,
11300 NE 2nd Ave., Miami Shores, FL 33161, USA}
\author{Alessandro Mirizzi}
\affiliation
{II Institut f\"ur Theoretische Physik, Universit\"at Hamburg, Luruper Chaussee 149, 22761 Hamburg, Germany} 
\author{Oscar Straniero}
\affiliation{INAF, Osservatorio Astronomico di Collurania, 64100 Teramo, Italy}
%

%*************************************   Abstract   **********************************************%
%\date{11 June}

\begin{abstract}
We derive a strong bound on the axion-photon coupling $g_{a\gamma}$ from the analysis of a sample of
39 Galactic Globular Clusters.
As recognized long ago, the $ R $ parameter, i.e. the number ratio of stars in horizontal over red giant branch of  old 
stellar clusters, would be reduced by the axion production from photon conversions  occurring in stellar cores. In this regard we have compared the measured $R$ with
 state-of-the-art stellar models obtained under different assumptions for 
$g_{a\gamma}$. We show  that
the estimated value of $g_{a\gamma}$ substantially 
depends on the adopted He mass fraction Y, an effect often neglected in previous investigations.   
 Taking as benchmark for our study 
the most recent determinations of the He abundance in  H {\small II} regions with O/H in the same range of the Galactic Globular Clusters,  
we obtain an upper bound $g_{a\gamma}<0.66\times 10^{-10}$ 
GeV$^{-1}$ at 95$\%$ confidence level. 
This result significantly improves  the constraints from previous analyses and is currently the strongest limit on the axion-photon coupling in a wide mass range.
\end{abstract}

%*************************************************************************************************%


\maketitle

\section{Introduction}---
Axions are low-mass pseudoscalar particles, somewhat similar to neutral pions.
Originally, they were introduced to explain the absence of CP violation in the strong interactions (QCD) \cite{Peccei:1977hh,Peccei:1977ur,Weinberg:1977ma,Wilczek:1977pj},
a long-standing puzzle in particle physics known as the strong CP problem.
Later on, it was also realized that the existence of such particles could account for most or all of the dark matter 
in the Universe.
Specifically, axions with masses in the $10~\mu{\rm eV}$  region would be cold dark matter candidates~\cite{Kawasaki:2013ae,Sikivie:2006ni,DiValentino:2014zna} 
while for  $m_a\agt60~{\rm meV}$ they would attain thermal equilibrium at the QCD phase transition 
or later~\cite{Turner:1986tb,Masso:2002np}, contributing to the cosmic radiation density and, subsequently
to the cosmic hot dark matter along with massive neutrinos~\cite{Archidiacono:2013cha}.

A generic property of axions is their two-photon coupling, specified by the Lagrangian
${\mathcal L}_{a\gamma}= g_{a\gamma} \textbf{E}\cdot \textbf{B}$, where
$g_{a\gamma} = 2\times 10^{-10}\,\ \rm{GeV}^{-1} \zeta \, (m_{a}/1\mbox{ eV}) $ and
$\zeta$ is a model dependent parameter of order one in many axion models.
This relation defines the ``axion line''
in the $m_a-g_{a\gamma}$ plane (see, e.g.,~\cite{Kim:2008hd}).
However,
in recent years, a considerable attention has been devoted to the so-called Axion-Like-Particles (ALPs)
which couple to photons, but do not satisfy the mass-coupling relation defined above for the QCD axions. 
Such light pseudoscalar particles emerge naturally in various extensions of the Standard Model 
(see, e.g.,~\cite{Jaeckel:2010ni}) and are phenomenologically motivated by a series of unexplained astrophysical 
observations. Among these, the seeming transparency of the universe to Very High-Energy gamma-rays~\cite{De Angelis:2007dy}, 
the larger than expected white dwarf cooling rates~\cite{Isern:2008nt}, and the quest for dark matter candidates 
(see~\cite{Sikivie:2009qn,Sikivie:2010bq,Carosi:2013rla} and references therein). 

As pointed out in a seminal paper by Pierre Sikivie~\cite{Sikivie:1983ip}, 
the two-photon coupling $a \gamma \gamma$ allows for efficient experimental searches of axions and ALPs.
 Indeed, in the presence of an external magnetic field, the $a \gamma \gamma$ coupling  leads to the phenomenon of photon-axion mixing~\cite{Raffelt:1987im}.
This  mechanism is the basis for \emph{ direct searches} of axions  in light-shining-through-the-wall experiments (see e.g.~\cite{Ehret:2010mh}) and 
axion dark matter in micro-wave cavity experiments (see e.g.~the ADMX experiment~\cite{Duffy:2006aa}).
Furthermore, the $g_{a\gamma}$ vertex would also allow for a production of axions via Primakoff process in 
stellar plasma~\cite{Raffelt:1985nk}. 
The predicted solar axion spectrum
 is  currently
searched by the CERN Axion Solar Telescope (CAST)~\cite{Andriamonje:2007ew}, 
looking for conversions into X-rays of solar axions in a dipole magnet directed towards the
sun. 
CAST searches 
with vacuum inside the magnet bores achieved a limit of
$g_{a\gamma} \lesssim 0.88 \times 10^{-10}$~GeV$^{-1}$ for 
$m_a \lesssim 0.02$~eV~\cite{Andriamonje:2007ew}, an excellent constraint for very light ALPs. 
 For  realistic QCD axions, CAST has explored the mass range up to $1.17$~eV, providing the bound 
$g_{a\gamma} \lesssim 2.3-3.3 \times 10^{-10}$~GeV$^{-1}$
 at 95~$\%$ CL, by using $^4$He~\cite{Arik:2008mq} and 
 $^3$He~\cite{Arik:2011rx,Arik:2013nya} as buffer gas.

The Primakoff process induced by the photon-axion coupling would also allow for \emph{indirect
axion searches}, via effects on stellar evolution. In this context,
additional constraints on the axion-photon coupling have been obtained from astronomical observations of
helium burning low and intermediate mass stars~\cite{Raffelt:1987yu,Raffelt:1996wa,Raffelt:2006cw,Friedland:2012hj}.
A recent analysis showed that a sufficiently large axion emission would affect the very existence of Cepheids variables in the mass range $M\sim 8-12$~$M_{\odot}$, providing the bound $g_{a\gamma} < 0.8 \times 10^{-10}$~GeV$^{-1}$~\cite{Friedland:2012hj}.
 On the other hand, photometric studies of Globular Cluster (GC) stars 
provided the long-standing strong  bound $g_{a \gamma}\lesssim 10^{-10}$~GeV$^{-1}$ 
for an axion mass lower than about 10 keV~\cite{Raffelt:1987yu,Raffelt:1996wa,Raffelt:2006cw}.

Globular Clusters  are gravitationally bound systems of stars populating the Galactic Halo.
They are among the oldest objects in the Milky Way. Hence only low mass stars (M$\lesssim 0.85$~M$_\odot$)
are still alive and, therefore, observable. A typical CG harbors a few millions  stars, so that 
the various evolutionary phases are well populated and distinguished from each other. In particular,
one can easily locate
the main sequence, corresponding to the core H burning phase, the red giant branch (RGB), during which the 
stellar luminosity is supported by the  H burning shell, and the horizontal branch (HB), corresponding to the 
core He burning phase. The number of stars observed in a particular evolutionary phase is proportional to the
corresponding lifetime, which is determined by the efficiency of all the relevant sources and sinks of energy.    
As early recognized,  axions coupled to photons would significantly reduce the lifetime of stars in the 
HB, while producing negligible changes on the  RGB evolution~\cite{Raffelt:1987yu}.%
~\footnote{The RGB phase has been recently exploited to set  a bound on the neutrino dipole moment and on the axion-electron coupling~\cite{Viaux:2013lha}.}
 Therefore,  
$g_{a\gamma}$ can be constrained by measurements of the $ R $ parameter,  
$R= {N_{\rm HB}}/{N_{\rm RGB}}$, 
which compares the numbers of stars in the HB  ($N_{\rm HB}$) and in the upper portion of the RGB
($N_{\rm RGB}$).

The previous analyses were based on the assumption that the measured $R$ parameter is well
reproduced, within 30\%, by extant models of GC stars, without including axion cooling.
Although it was recognized long ago (see, e.g.,~\cite{iben,rood,buzzoni,raffeltry}) that the $R$ parameter is sensitive to the helium
 mass fraction Y, which mainly affects the number of RGB stars,  in the context of the axion bounds this dependence 
 has so far been neglected. 
Indeed, even a considerable
decrease of the HB lifetime caused by a large value of $g_{a\gamma}$ could be compensated by a suitable 
increase of the assumed He content.
Because of this degeneracy, a proper evaluation of the axion constraints from the $R$ parameter relies on our 
knowledge of the He abundance in the GCs.
He abundance measurements  are particularly difficult for 
Globular Clusters stars. However, since they are among the first stars appeared in the Universe, it is commonly
assumed that the original He content of Galactic GCs practically coincides with the primordial one (Y$_p$). 
At this regard,
in the last 20 years the estimation of Y$_p$ has improved significantly, changing from
$\sim 0.23$~\cite{Olive:1994fe} to $\sim 0.25$~\cite{Izotov:2013waa}.
Furthermore, the large amount of new photometric studies of GCs accumulated 
over the last 20 years by exploiting Earth and space based telescopes 
allows a more accurate determination of the $R$ parameter~\cite{Salaris:2004xd}.

In light of these improvements and of the great importance of the GC bound for the current experimental 
efforts, we provide here a new analysis of this 
astrophysical constraint including, for the first time, the effects of the helium mass fraction.
Our result, $g_{a\gamma}<0.66\times 10^{-10}$ GeV$^{-1}$ at 95$\%$ confidence level, improves significantly the bound from the previous analyses and is currently the strongest constraint on the axion-photon coupling in a wide mass range.

\section{Analysis}---
Salaris \emph{et al.}~\cite{Salaris:2004xd}  reported measurements of the $R$ parameter for a sample of 57 galactic clusters. 
As discussed below, for the star's total metal abundance [M/H]$<-1.1$~\footnote{Here, we are using the standard spectroscopic notation for the 
relative abundances, [M/H]$=\log_{10}(\mathrm{Z}/\mathrm{X})-\log_{10}(\mathrm{Z}/\mathrm{X})_\odot$, where X 
is the hydrogen mass fraction and Z is the total mass fraction of all the elements except H and He, i.e., Z=1-X-Y.}
the $R$ parameter is practically independent of the cluster age and metallicity. 
At larger metallicity, however, the so-called 
RGB ``bump''~\footnote{The ``bump'' is an intrinsic feature  appearing as a peak
in the differential luminosity function of GCs. 
It originates when the H-burning shell crosses the chemical discontinuity left over by the convective envelope 
soon after the first dredge-up, slowing down the evolutionary timescale.}
is too faint to enter into the RGB star count and, 
in turn, the resulting $R$ is definitely larger. Therefore, in our analysis we considered only the 39 
clusters with [M/H]$<-1.1$, for which we obtain a weighted average
$R_{\rm ave}=1.39\pm 0.03$, and assumed  that all the stars of the  39 clusters sample share 
the same original He abundance. 
The small statistical error (about 2\%) supports this hypothesis. 

It has been suggested that some GCs may harbor He enhanced stellar populations (see~\cite{Gratton:2012ap}).
Indeed, the presence of He-rich stars would lead to a certain overestimation of the $R$ parameter. 
However, He enhanced stars would be less massive than coeval stars with primordial He content,
so that they would be located in the bluer part of the HB. We have tested this 
possibility by restricting the cluster sample, considering  only 18 clusters 
whose HB is not dominated by blue stars~\footnote{The selection has been made by including only clusters 
with $(n_B-n_V)/(n_B+n_V+n_R)<0.8$, where
$n_B$, $n_V$ and $n_R$ represent the number of HB stars
bluer than the RR Lyrae instability strip, within the strip
and redder than the strip, respectively~\cite{Lee:1994pd}.}.
The new weighted average $R_{\rm ave}=1.39\pm 0.04$, practically coincides
with the one obtained for the whole sample, thus supporting the 
usual assumption  that the bulk of the stars in our GC sample shares
a unique He abundance.

 Axions or ALPs with mass below a few keV could be produced in stellar interiors via the Primakoff process -- 
the conversion of a photon into an axion  in the fluctuating 
electric field of nuclei and electrons in the stellar plasma~\cite{Raffelt:1985nk}.
 Being weakly interacting, axions would efficiently carry energy outside the star, much like neutrinos do, providing an 
effective cooling mechanism.
In the following, we will neglect other possible couplings of axions with nucleons and electrons, since 
these are rather model dependent (see, e.g.,~\cite{Kim:2008hd}). If present, also these interactions would
contribute to the energy-loss. In this respect, our limit on $g_{a\gamma}$  should be considered conservative.

%
%Assuming a non-degenerate plasma and ignoring the plasma frequency, the Primakoff production rate is~\cite{Raffelt:1985nk}
%\begin{equation}
%\epsilon_{a}= Z(\xi^{2})\frac{g_{a\gamma}^{2}}{4\pi^{2}} \frac{T^{7}}{ \rho} 
%\simeq 28 \frac{\mbox{erg}}{\mbox{g}~\mbox{s} } Z(\xi^{2}) g_{10}^{2} T_{8}^{7} \rho_{4}^{-1}, 
%\label{eq:loss}
%\end{equation}
%where $g_{10}\equiv g_{a\gamma}/(10^{-10}$~GeV$^{-1}$), 
%$\rho_4\equiv\rho/(10^4$~g/cm$^{3})$, $T_8 \equiv T/10^8 \rm K$ and 
%$Z(\xi^{2})$ is a function (see, e.g., \cite{Raffelt:1996wa}), generally ${\cal O}(1)$ for relevant stellar conditions,
%of $\xi^{2}\equiv(\kappa_{S}/2T)^{2}$, with $\kappa_{S}$ being the Debye-Huckel screening wavenumber.

In order to asses the axion effects on stellar evolution and derive a bound on  $ g_{a\gamma} $, 
we have computed several evolutionary sequences of stellar models, from the pre-main-sequence to the 
asymptotic giant branch,  with different initial mass (M), RGB mass loss rate, 
metallicity (Z), helium mass fraction (Y) and axion coupling
 ($g_{a\gamma}$).
 The models were computed 
by means of  FUNS  (FUll Network Stellar evolution), an hydrostatic 1D stellar evolution 
code~\cite{Straniero:2005hc,Luciano:2013rya, straniero2014}.
Axion effects have been introduced as an additional energy sink following the procedure in~\cite{Raffelt:1987yu}
which includes the effects of electron degeneracy and of non-zero plasma frequency, relevant for the evolution
during the RGB phase.

Besides axion induced effects, proportional to $ g_{a\gamma}^2 $, 
%as expected from Eq. (\ref{eq:loss}), 
variations of $R$ may be caused by changes of the parameters
characterizing the cluster, such as age, metallicity or He content. 
Our numerical analysis shows negligible variations of $R$ for initial stellar 
masses in the range $0.82\le$ M/M$_\odot \le 0.84$ and metallicities in 
$0.0002\le$ Z $\le 0.001$, which correspond to cluster ages 
between 11.1 and 13.3 Gyr and $-1.9\le$ [M/H] $\le -1.1$, respectively. 
On the other hand, we find a linear dependence of $R$ on the He mass fraction 
of the cluster. 
The relation
\begin{eqnarray}\label{eq:erreth}
R_{\rm th} (g_{a \gamma}, Y)=6.26\, Y-0.41\, g_{10}^2 -0.12\,,
\end{eqnarray}
describes very well our numerical results and shows the mentioned degeneracy between Y and $g_{a \gamma}$.
Evidently, an accurate determination of the He mass 
fraction in GCs is necessary to appropriately constrain the axion-photon coupling.
 As mentioned above measurements of helium abundance in GC stars are challenging. 
Indeed, ultraviolet data are needed to perform He
abundance analysis in stars, a spectroscopic window not achievable from Earth. In addition, 
convection, rotational induced mixings and other secular phenomena, such as gravitational settling, modify the He abundance in the atmospheres of these stars. 
For this reason, the primordial He is often adopted for GC stars.
Actually, Y$_p$ represents a lower bound for the GC He mass fraction.
%Indirect estimations of Y$_p$ can be derived from Big Bang Nucleosynthesis
%(BBN) calculations~\cite{Iocco:2008va}. 
%In the Standard BBN scenario, the Planck experiment predicted Y$_p$=0.24725 $\pm 0.00032$ 
%at 68~\% confidence level, for the measured value of 
%baryon density~\cite{Ade:2013zuv}.
%However, this result significantly depends on the expansion rate and,
%in
%turn, on the assumed number of neutrinos $N_{\rm eff}$~\cite{Iocco:2008va}.
%Indeed, the Planck analysis finds Y$_p$=0.254$^{+0.041}_{-0.033}$ at 68~\% confidence level~\cite{Ade:2013zuv},
%when $N_{\rm eff}$ is left free. 
For our purpose, we prefer to use direct measurements of Y in low metallicity environments which may be considered 
representative of the chemical composition of the early Galaxy.  
In this context, optical spectra of low-metallicity H {\small II} regions show several He {\small I} lines which allow
a quite accurate He abundance determination.                                                                                                                                                                                                                                                                                                                                                                                                                                                                                                                                                                                                                                                                                                                                                                                                                                                                                                                                                                                                                                                                                                                                                                                                                             
The most recent independent studies  of low-metallicity  H {\small II} regions are those published by 
Izotov \emph{et al.}~\cite{Izotov:2013waa} and by Aver \emph{et al.}~\cite{Aver:2013wba}. 
These two groups use very similar procedures and tools, but different datasets. In particular, 
Aver et al. use high accuracy spectra of 16 Blue Compact Dwarfs Galaxies with $1.5 < $O/H$(\times10^5) < 13$.
Note that this range of O/H is approximately the same of the 39 GCs we have used to derive the $ R $ parameter. 
The 111  H {\small II} regions used by Izotov \emph{et al.}~\cite{Izotov:2013waa} extend to larger metallicity, even though most of them have   O/H is in the same range as Aver \emph{et al.}~\cite{Aver:2013wba}.   
In spite of the different datasets, the resulting weighted average values for the He abundance are very similar,
namely: Y=0.2535$ \pm $0.0036 and  0.255$ \pm $0.003 for Aver \emph{et al.}~\cite{Aver:2013wba} and  Izotov \emph{et al.}~\cite{Izotov:2013waa}, respectively 
\footnote{These average values shouldn't  be confused with the extrapolated values at 0 metallicity calculated 
by both groups, which represent an estimation of the primordial He.}.                                                                                                                                                                                                                                                                                                                                                                                                                                                                                                                                                                                                                                                                                                                                                                                                                                                                                                                                                                                                                                                                                                                                                                                                                                                                                                                                                  
Since the result obtained by Izotov \emph{et al.} could be slightly higher, because of the few high Z H {\small II} regions included in their 
dataset,  in the following we will use the weighted average value reported by Aver \emph{et al.}~\cite{Aver:2013wba} for the same metallicity range of the
39 GCs of our sample.

 %%%%%%%%%%%%%%%%%%%%%%%%%%%%%%%%%%%%%%%%%%%%%%%%%%%%%%%%%%%%%%%%%%%%%%%%%%%%%%%%%%%%%%%%%%%%%%%%%%%%%%55
\begin{figure}[bt]
\includegraphics[angle=0,width=1.\columnwidth]{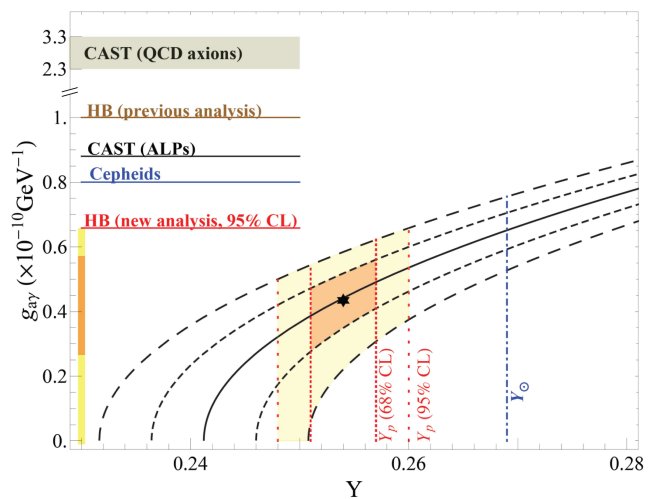}
\vspace{-0.5cm}  
\caption{$R$ parameter constraints to Y and $g_{a\gamma}$. 
The vertical lines indicate respectively the $1\sigma$ (short-dotted curves) and $2\sigma$ (long-dotted curves)
of Y. The dot-dashed vertical line indicate the preferred  value of Y$_{\odot}$.
The other bent curves correspond to the determination of $g_{a\gamma}$ as function of 
Y from $R_{\rm th}$ [Eq.~(\ref{eq:erreth})]. Specifically,
the continuous curve corresponds to  $R_{\rm th}=R_{\rm ave}$, while the short and long-dashed lines indicate, 
respectively, the 1$\sigma$ and the 2$\sigma$ ranges.
The star represents the best 
  fits for Y$=0.254$. The  shaded area  delimits the combined 68\% CL (dark)
and 95\% CL (light) for Y and $R_{\rm th}$.
  The vertical rectangles indicate the 68\% CL (dark) and 95\% CL (light) for $g_{a\gamma}$.
  Previous bounds from HB lifetime~\cite{Raffelt:1987yu}, from the
Cepheids observation~\cite{Friedland:2012hj},  from CAST for light ALPs~\cite{Arik:2011rx,Arik:2013nya}
and for QCD axions~\cite{Andriamonje:2007ew} are also shown.}
  \label{fig:G10VsY}
\end{figure}   
%%%%%%%%%%%%%%%%%%%%%%%%%%%%%%%%%%%%%%%%%%%%%%%%%%%%%%%%%%%%%%%%%%%%%%%%%%%%%%%%%%%%%%%%%%%%%5

%%%%%%%%%%%%%%%%%%%%%%%%%%%%%%%%%%%%%%%%%%%%%%%%%%%%%%%%%%%%%%
\section{The new bound for the axion-photon coupling}---
%%%%%%%%%%%%%%%%%%%%%%%%%%%%%%%%%%%%%%%%%%%%%%%%%%%%%%%%%%%%%%%%%%%
In order to constrain the axion-photon coupling,
we compare the average value of  $ R $ ($R_{\rm ave}$)  with 
the theoretical prediction ($R_{\rm th}$). Assuming
that the $R$  measurements are distributed as Gaussian variables, one can determine
confidence levels for the different quantities. 
Our results are shown in Fig.~\ref{fig:G10VsY}. 
The vertical lines indicate, respectively, 68\% CL (short-dotted curves) and 95 \% CL (long-dashed curves)
uncertainties of Y. 
The other bent curves correspond to the determination of $g_{a\gamma}$ as function of 
Y from $R_{\rm th}$ [Eq.~(\ref{eq:erreth})]. In particular,
the solid black curve has been obtained  with $R_{\rm th}=R_{\rm ave}$,  while 
the short-dashed  and the long-dashed black lines indicate,  respectively, the  $1\sigma$ and the $2\sigma$ ranges.

%Given the anti-correlation between  Y and $g_{a \gamma}$, 
%a small axion-photon coupling improves the agreement
%between models and observations. 
Combining the confidence levels of Y and 
$R_{\rm th}$,  we find 
\begin{eqnarray}\label{Eq:1sigma}
g_{a \gamma} = 0.45_{-0.16}^{+0.12} \times 10^{-10}~{\rm GeV}^{-1} \,\  \,\ (68 \% \,\ \textrm{CL})  \,\ ,
\end{eqnarray}
(the best-fit point is indicated with a star in Fig.~\ref{fig:G10VsY})
while  
 %......................................................
\begin{equation} 
 g_{a\gamma}<0.66\times 10^{-10}\,\ \textrm{GeV}^{-1} \,\  \,\ (95 \% \,\ \textrm{CL})  \,\ .
 \label{Eq:2sigma}
\end{equation}
Note that in the standard physics scenario, $g_{a\gamma}=0$, 
we find Y$=0.241 \pm 0.005$ which is   
compatible with the measured Y at $2\sigma$.
%Given the slight discrepancy between the Y measured in low metallicity HII regions and the one inferred from $R_{\rm th}$,
%one could be tempted to  interpret our result as a hint for axions. However,  its low statistical
%significance and the  possible
%uncontrolled systematic uncertainties in the $R$ measurement suggest  to thrust only
% the  $2\sigma$ upper bound, coming from the combinations of the confidence levels
%of $R_{\rm th}$ and Y, and shown by the light shaded area in the Figure. 
%Of course,  possible improvements of the observations may eventually lead to an increased confidence in a new physics effect. 

As we have shown, the largest source of systematic error is the adopted helium mass 
fraction. Certainly the primordial He provides a lower bound to the GC He. 
 For istance, by taking the latest SBBN prediction, as obtained after the Plank results~\cite{Ade:2013zuv},
we would obtain a more stringent constraint for the axion-photon coupling, i.e.,
$g_{a\gamma}<0.50\times 10^{-10}\,\ \textrm{GeV}^{-1} \,\  \,\ (95 \% \,\ \textrm{CL})  \,$. 
%One may argue that some chemical evolution
%occurred before the formation of the GC system so that their He content
%may be larger than the one we have assumed.
 On the other hand, the He
content of the solar system provides a very conservative upper bound for
the GC He. The He abundance in the early 
solar system is an input parameter of Standard Solar Models  and its value is  
mostly constrained by the present solar system age, as derived by means of radioactive dating 
techniques of terrestrial and meteoritic materials. Piersanti \emph{et al.}~\cite{Piersanti:2006zh} 
 found Y$_\odot =0.269$ (vertical dot-dashed line 
in Fig.~1) in good agreement with other extant 
Standard Solar Models.~\footnote{Serenelli \emph{et al.}~\cite{Serenelli:2010fk}, by adopting the helioseismic determination 
of the present-day solar surface He abundance, found a slightly larger
value, i.e. Y$_\odot =0.278$.} 
By using this solar He mass fraction, 
we find a higher upper bound, namely
 $g_{a \gamma} < 0.76 \times 10^{-10}$~GeV$^{-1}$ (95\% CL). 
 However,  this is an overly conservative 
assumption which would imply that no chemical evolution  occurred during
the 8 Gyr elapsed between the GC and the solar system formation, in
contrast with many well-known astronomical evidences. 
%Therefore,
%the result obtained by adopting the Izotov \emph{et al.}~\cite{Izotov:2013waa} 
%result for the primordial He is definitely more
%realistic.

\begin{table}
\caption{axion-photon coupling bounds}\label{tab1}
\centering
\begin{tabular}{c c c c c}
\hline
       &     &   R    &   Y   &  $g_{10}$  \\ 
\hline
 bounds from low-Z II regions  & up 95\% & 1.33 & 0.260 & 0.66 \\
 -               & up 68\% & 1.36 & 0.257 & 0.57 \\
 -               & central value & 1.39 & 0.254 & 0.45 \\ 
 -               & low 68\% & 1.42 & 0.251 & 0.29 \\
 -               & low 95\% & 1.45 & 0.248 & 0.00 \\
& & & & \\
 bounds from SBBN  & up 95\% & 1.33 & 0.2478 & 0.50 \\
 -               & up 68\% & 1.36 & 0.2475 & 0.42 \\
 -               & central value & 1.39 & 0.2472 & 0.31 \\ 
 -               & low 68\% & 1.42 & 0.2469 & 0.15 \\
 -               & low 95\% & 1.45 & 0.2466 & 0.00 \\
  & & & & \\
 bounds from Y$_\odot$  & up 95\% & 1.33 & 0.269 & 0.76 \\
 -               & up 68\% & 1.36 & 0.269 & 0.71 \\
\hline
\end{tabular}
\end{table}

In Table~I we  summarize the various bounds obtained under the different
assumptions on Y. 
 Obviously, our analysis relies on the reliability of the adopted stellar models 
 of RGB and HB stars. In the Appendix
 we will give a short summary of the state-of-the-art. A detailed study of the relevant uncertainties will be extensively presented in a forthcoming paper.
 
\section{Discussion and Conclusions}---
We have obtained a new and more stringent  bound on the axion-photon coupling constant
$g_{a\gamma}$ 
 from an updated analysis of the $R$ parameter in 39 Galactic GCs.
 %In addition, our analysis underlines an anti-correlation between 
 %$g_{a\gamma}$ and 
%the adopted He mass fraction $Y$, an effect neglected in  
%previous investigations.
%Finally, we  point out a slight discrepancy (at $1\sigma$) between the predicted helium fraction and the one inferred from 
%$R$, which would be mitigated  by a non-zero axion-photon coupling as given in Eq.~(\ref{Eq:1sigma}).
Our constrain, given in Eq.~(\ref{Eq:2sigma}),
represents the strongest limit on $g_{a\gamma}$ for QCD axions in a wide
mass range.
Only in the case of cold dark matter axions there is a stronger constrain,
 $g_{a\gamma}\lesssim 10^{-15}$ GeV$^{-1}$ from ADMX, and only for a narrow range around $m_a\sim 1$~$\mu$eV~\cite{Asztalos:2009yp}.
As evident from Fig.~1, our result 
 improves the 
previous long-standing bound from GCs~\cite{Raffelt:1987yu},
$g_{a\gamma}\lesssim  10^{-10}$ GeV$^{-1}$ and the more recent one from Cepheid stars,
$g_{a\gamma}\lesssim  0.8\times 10^{-10}$ GeV$^{-1}$~\cite{Friedland:2012hj}. 
Moreover, it is a factor $\sim 4$ better than the current experimental bound
on QCD axions from the CAST experiment (see Fig.~1).
This is also the strongest constraint 
for generic ALPs, except 
 in the extremely low mass region $m_a \lesssim 10^{- 10}$~eV. 
There, a more stringent limit $g_{a\gamma}\lesssim  10^{-11}$~GeV$^{-1}$~\cite{Brockway:1996yr} or even $g_{a\gamma} \lesssim 3\times 10^{-12}$~GeV$^{-1}$~\cite{Grifols:1996id} has been derived from the absence of $\gamma$-rays from SN~1987A.

Ultra-light ALPs with such a small coupling would play an important role in astrophysics. 
A particularly intriguing hint for
these particles has been recently suggested by Very High-Energy gamma-ray 
experiments~\cite{De Angelis:2007dy}, even though
 this problem has been also analyzed using more conventional physics (see, e.g.,~\cite{Essey:2009zg,Essey:2009ju}).
Indeed, photon-axion conversions in large-scale cosmic magnetic fields would reduce the opacity of the universe
to TeV photons, explaining the anomalous spectral hardening
 found in the Very High-Energy gamma-ray spectra~\cite{Horns:2012fx}. 
In particular, for realistic models of the cosmic magnetic field, this scenario would require
$g_{a \gamma} \gtrsim 0.2 \times 10^{-10}$ GeV$^{-1}$ and $m_a \lesssim 10^{-7}$~eV~\cite{Meyer:2013pny}.

Remarkably, the coupling ranges discussed in this letter are accessible by new independent laboratory searches, such as 
the planned upgrade of the photon regeneration experiment ALPS at DESY~\cite{Ehret:2010mh,Bahre:2013ywa} and  the next generation solar axion detector IAXO (International Axion Observatory)~\cite{Irastorza:2011gs}.
This confirms, once again, the nice synergy between astrophysical arguments and laboratory searches to corner axions and
 axion-like particles.

%%%%%%%%%%%%%%%%%%%%%%%%%%%%%%%%%%%%%%%%%%%%%%%%%%%%%%%%%%%%%%%%%%%%%%%%%%%%%%%%%%%%%%%%%%%%%%5

$ $

We thank  Daniele Montanino, Maurizio Paolillo and Francesco Villante
 for interesting discussions during the initial phases of this work.
We also thank Georg Raffelt and Pasquale Dario Serpico for valuable comments on the manuscript.  
 A.A., I.D., and O.S. acknowledge
 financial support from the Spanish Ministry of Economy and
Competitiveness project AYA2011-22460.
The work of  A.M. was supported by the German Science Foundation (DFG) within the Collaborative Research Center 676 ``Particles, Strings and the Early Universe''.

%%%%%%%%%%%%%%%%%%%%%%%%%%%%%%%%55
\section{Appendix A: Astrophysical uncertainties}---
 In the context of the present study, 
energy sources and sinks are the most important phenomena to take under control, 
because they directly affect stellar lifetimes. 
During the RGB phase, the nuclear energy production is regulated by the 
$^{14}$N$(p,\gamma)^{16}$O, which acts as a bottleneck of the shell H burning.
After a dedicated experimental investigation made by the LUNA
 collaboration~\cite{Lemut:2006va}, a very accurate measurement of the astrophysical factor for  
$^{14}$N$(p,\gamma)^{16}$O is now available down to about 70 keV. 
The uncertainty of the reaction rate between 50 and 100 MK is
lower than 10\%. 
Concerning the HB phase, the $^{4}$He$(2\alpha,\gamma)^{12}$C~ and the 
$^{12}$C($\alpha,\gamma)^{16}$O~ reactions compete during the 
core H burning. For the triple$-\alpha$ reaction rate the uncertainty 
at the temperature of the core-He burning is expected 
to be lower than 10\%\cite{Angulo:1999zz,Fynbo:2005vf}.  
Definitely larger uncertainties affects the $^{12}$C($\alpha,\gamma)^{16}$O~ reaction. 
A recent R-matrix analysis, which includes all the available direct
 measurements~\cite{Schurmann:2012zz}, confirms previous finding~\cite{Kunz:2001zz}, 
but reduces substantially the error bar. 
At E=300 keV they find $S(300)=161\pm 19_{\,\ \rm stat}(^{+8}_{\rm -2\,\ sys})$.
For the models  presented here we have used the  rate of~\cite{Kunz:2001zz}. Note, however, that
the $^{12}$C($\alpha,\gamma)^{16}$O~ reaction 
only contributes to the energy production during the last 10-15\% of the HB lifetime.
Its rate mostly affects the final amount of C and O left in the core at the
end of the He burning~\cite{Straniero:2003ax}. Asteroseismic studies of pulsating white dwarf
may be used to evaluate the internal chemical stratification and, in turn, to constrain core
 He burning models~\cite{Metcalfe:2003ti}. As shown in~\cite{Straniero:2003ax}, 
the C/O near the WD center also depends 
on the extension of the convective core that develops during the He burning. 
In particular, it was found that by means of the same treatment of convection we have adopted  
in the present study for HB models, the best agreement with the C/O measured in 
pulsating WDs is obtained when the $^{12}$C($\alpha,\gamma)^{16}$O reaction rate suggested~\cite{Kunz:2001zz} 
and confirmed by~\cite{Schurmann:2012zz}
 is adopted. In other words, the 
WD constraint implies that a variation of the modeled convective boundary
requires a corresponding compensative change of the $^{12}$C($\alpha,\gamma)^{16}$O rate. 
More relevant for the present analysis is that this compensation 
limits the possible variations of the HB lifetime, the quantity explicitly 
used in the R parameter calculation.  
Concerning energy sinks, plasma neutrino loss plays a 
fundamental role in the RGB evolution. We adopt the latest neutrino rate reported 
by Esposito \emph{et al.}~\cite{Esposito:2003wv}, which is in 
excellent agreement with previous independent calculations~\cite{Haft:1993jt,itoh}.
 
 $ $

%%%%%%%%%%%%%%%%%%%%%%%%%%%%%%%%%%%%%%%%%%%%%%%%%%%%%%%%%%%%%%%%%%%%%%
%\section*{References} %%%%%%%%%%%%%%%%%%%%%%%%%%%%%%%%%%%%%%%%%%%%%%%%
%%%%%%%%%%%%%%%%%%%%%%%%%%%%%%%%%%%%%%%%%%%%%%%%%%%%%%%%%%%%%%%%%%%%%%

%*************************************   Bibliography   ******************************************%


\vspace{-0.5cm}   

\begin{thebibliography}{99} 

%%% Axion papers %%%%

%\cite{Peccei:1977hh}
\bibitem{Peccei:1977hh} 
  R.~D.~Peccei and H.~R.~Quinn,
  %``CP Conservation in the Presence of Instantons,''
  Phys.\ Rev.\ Lett.\  {\bf 38}, 1440 (1977).
  %%CITATION = PRLTA,38,1440;%%
  %2808 citations counted in INSPIRE as of 14 Jun 2014

%\cite{Peccei:1977ur}
\bibitem{Peccei:1977ur} 
  R.~D.~Peccei and H.~R.~Quinn,
  %``Constraints Imposed by CP Conservation in the Presence of Instantons,''
  Phys.\ Rev.\ D {\bf 16}, 1791 (1977).
  %%CITATION = PHRVA,D16,1791;%%
  %1711 citations counted in INSPIRE as of 14 Jun 2014
  
%\cite{Weinberg:1977ma}
\bibitem{Weinberg:1977ma} 
S.~Weinberg,
%``A New Light Boson?,''
Phys.\ Rev.\ Lett.\  {\bf 40}, 223 (1978).
%%CITATION = PRLTA,40,223;%%
%1989 citations counted in INSPIRE as of 14 Jun 2014
    
%\cite{Wilczek:1977pj}
\bibitem{Wilczek:1977pj} 
  F.~Wilczek,
  %``Problem of Strong p and t Invariance in the Presence of Instantons,''
  Phys.\ Rev.\ Lett.\  {\bf 40}, 279 (1978).
  %%CITATION = PRLTA,40,279;%%
  %1920 citations counted in INSPIRE as of 14 Jun 2014


%\cite{Sikivie:2006ni}
\bibitem{Sikivie:2006ni}
  P.~Sikivie,
%  ``Axion cosmology,''
  Lect.\ Notes Phys.\  {\bf 741}, 19 (2008) 
  [astro-ph/0610440].


%\cite{Kawasaki:2013ae}
\bibitem{Kawasaki:2013ae}
  M.~Kawasaki and K.~Nakayama,
 % ``Axions: Theory and cosmological role,''
  Ann.\ Rev.\ Nucl.\ Part.\ Sci.\ {\bf 63}, 69 (2013) 
  [arXiv:1301.1123].
  %%CITATION = ARXIV:1301.1123;%%
  %5 citations counted in INSPIRE as of 23 Apr 2013


\bibitem{DiValentino:2014zna} 
  E.~Di Valentino, E.~Giusarma, M.~Lattanzi, A.~Melchiorri and O.~Mena,
  %``Axion cold dark matter: status after Planck and BICEP2,''
  arXiv:1405.1860 [astro-ph.CO].


\bibitem{Turner:1986tb}
  M.~S.~Turner,
%  ``Thermal production of not so invisible axions in
%  the early universe,''
  Phys.\ Rev.\ Lett.\  {\bf 59}, 2489 (1987);
  Erratum ibid.\ {\bf 60}, 1101 (1988).
  %%CITATION = PRLTA,59,2489;%%

\bibitem{Masso:2002np}
  E.~Mass\'o, F.~Rota and G.~Zsembinszki,
 % ``On axion thermalization in the early universe,''
  Phys.\ Rev.\ D {\bf 66}, 023004 (2002) 
  [hep-ph/0203221].


%\cite{Archidiacono:2013cha}
\bibitem{Archidiacono:2013cha} 
  M.~Archidiacono, S.~Hannestad, A.~Mirizzi, G.~Raffelt and Y.~Y.~Y.~Wong,
  %``Axion hot dark matter bounds after Planck,''
  JCAP {\bf 1310}, 020 (2013)
  [arXiv:1307.0615 [astro-ph.CO]].
  
  
%\cite{Kim:2008hd}
\bibitem{Kim:2008hd}
  J.~E.~Kim and G.~Carosi,
 % ``Axions and the strong CP problem,''
  Rev.\ Mod.\ Phys.\  {\bf 82}, 557 (2010) 
  [arXiv:0807.3125].
  %%CITATION = ARXIV:0807.3125;%%
  %159 citations counted in INSPIRE as of 23 Apr 2013



  
%\cite{Jaeckel:2010ni}
\bibitem{Jaeckel:2010ni}
  J.~Jaeckel and A.~Ringwald,
 % ``The Low-Energy Frontier of Particle Physics,''
  Ann.\ Rev.\ Nucl.\ Part.\ Sci.\  {\bf 60}, 405 (2010)
  [arXiv:1002.0329 [hep-ph]].
  %%CITATION = ARNUA,60,405;%%
  
  
  
    %\cite{De Angelis:2007dy}
\bibitem{De Angelis:2007dy}
  A.~De Angelis, M.~Roncadelli and O.~Mansutti,
  %``Evidence for a new light spin-zero boson from cosmological gamma-ray propagation?,''
  Phys.\ Rev.\ D {\bf 76} (2007) 121301
  [arXiv:0707.4312 [astro-ph]].
  %%CITATION = ARXIV:0707.4312;%%
  
 %\cite{Isern:2008nt}
\bibitem{Isern:2008nt} 
  J.~Isern, E.~Garcia-Berro, S.~Torres and S.~Catalan,
  %``Axions and the cooling of white dwarf stars,''
  Astrophys.\ J.\ Lett.\ {\bf 682}, L109 (2008)
  [arXiv:0806.2807 [astro-ph]].
  %%CITATION = ARXIV:0806.2807;%% 
  
  
  

%\cite{Sikivie:2009qn}
\bibitem{Sikivie:2009qn} 
  P.~Sikivie and Q.~Yang,
  %``Bose-Einstein Condensation of Dark Matter Axions,''
  Phys.\ Rev.\ Lett.\  {\bf 103}, 111301 (2009)
  [arXiv:0901.1106 [hep-ph]].
  %%CITATION = ARXIV:0901.1106;%%
  %102 citations counted in INSPIRE as of 18 Jun 2014

%\cite{Sikivie:2010bq}
\bibitem{Sikivie:2010bq} 
  P.~Sikivie,
  %``The emerging case for axion dark matter,''
  Phys.\ Lett.\ B {\bf 695}, 22 (2011)
  [arXiv:1003.2426 [astro-ph.GA]].
  %%CITATION = ARXIV:1003.2426;%%  
  
%\cite{Carosi:2013rla}
\bibitem{Carosi:2013rla} 
  G.~Carosi, A.~Friedland, M.~Giannotti, M.~J.~Pivovaroff, J.~Ruz and J.~K.~Vogel,
  %``Probing the axion-photon coupling: phenomenological and experimental perspectives. A snowmass white paper,''
  arXiv:1309.7035 [hep-ph].
  %%CITATION = ARXIV:1309.7035;%%
  %1 citations counted in INSPIRE as of 13 Jun 2014
  
  
   \bibitem{Sikivie:1983ip} 
  P.~Sikivie,
  %``Experimental Tests of the Invisible Axion,''
  Phys.\ Rev.\ Lett.\  {\bf 51}, 1415 (1983)
  [Erratum-ibid.\  {\bf 52}, 695 (1984)].



  
\bibitem{Raffelt:1987im}
  G.~Raffelt and L.~Stodolsky,
%  ``Mixing of the photon with low mass particles,''
  Phys.\ Rev.\ D {\bf 37}, 1237 (1988).
  %%CITATION = PHRVA,D37,1237;%%
  
  
%\cite{Ehret:2010mh}
\bibitem{Ehret:2010mh} 
  K.~Ehret, M.~Frede, S.~Ghazaryan, M.~Hildebrandt, E.~-A.~Knabbe, D.~Kracht, A.~Lindner and J.~List {\it et al.},
  %``New ALPS Results on Hidden-Sector Lightweights,''
  Phys.\ Lett.\ B {\bf 689}, 149 (2010)
  [arXiv:1004.1313 [hep-ex]].
  
  
 
%\cite{Duffy:2006aa}
\bibitem{Duffy:2006aa}
  L.~D.~Duffy {\it et al.},
%  ``A High Resolution Search for Dark-Matter Axions,''
  Phys.\ Rev.\  D {\bf 74}, 012006 (2006)
  [astro-ph/0603108].
  %%CITATION = PHRVA,D74,012006;%%
  
  
   
  %\cite{Raffelt:1985nk}
\bibitem{Raffelt:1985nk} 
  G.~G.~Raffelt,
  %``Astrophysical Axion Bounds Diminished By Screening Effects,''
  Phys.\ Rev.\ D {\bf 33}, 897 (1986). 
  
  
 \bibitem{Andriamonje:2007ew} 
  S.~Andriamonje {\it et al.}  [CAST Collaboration],
  %``An Improved limit on the axion-photon coupling from the CAST experiment,''
  JCAP {\bf 0704}, 010 (2007)
  [hep-ex/0702006].
  %%CITATION = HEP-EX/0702006;%% 
  
 



 
  
 \bibitem{Arik:2008mq} 
  E.~Arik {\it et al.}  [CAST Collaboration],
  %``Probing eV-scale axions with CAST,''
  JCAP {\bf 0902}, 008 (2009)
  [arXiv:0810.4482 [hep-ex]].
  %%CITATION = ARXIV:0810.4482;%% 
  



%\cite{Arik:2011rx}
\bibitem{Arik:2011rx} 
  S.~Aune {\it et al.}  [CAST Collaboration],
  %``CAST search for sub-eV mass solar axions with 3He buffer gas,''
  Phys.\ Rev.\ Lett.\  {\bf 107}, 261302 (2011)
  [arXiv:1106.3919 [hep-ex]].
  %%CITATION = ARXIV:1106.3919;%%
  
  
  %\cite{Arik:2013nya}
\bibitem{Arik:2013nya} 
  M.~Arik, S.~Aune, K.~Barth, A.~Belov, S.~Borghi, H.~Brauninger, G.~Cantatore and J.~M.~Carmona {\it et al.},
  %``CAST solar axion search with $^3$He buffer gas: Closing the hot dark matter gap,''
  Phys.\ Rev.\ Lett.\  {\bf 112}, 091302 (2014)
  [arXiv:1307.1985 [hep-ex]].
  %%CITATION = ARXIV:1307.1985;%%
  
   
  %\cite{Raffelt:1987yu}
\bibitem{Raffelt:1987yu} 
  G.~G.~Raffelt and D.~S.~P.~Dearborn,
  %``Bounds on Hadronic Axions From Stellar Evolution,''
  Phys.\ Rev.\ D {\bf 36}, 2211 (1987).
  %%CITATION = PHRVA,D36,2211;%% 
  
 %\cite{Raffelt:2006cw}
\bibitem{Raffelt:2006cw} 
  G.~G.~Raffelt,
  %``Astrophysical axion bounds,''
  Lect.\ Notes Phys.\  {\bf 741}, 51 (2008)
  [hep-ph/0611350].
  %%CITATION = HEP-PH/0611350;%% 

%\cite{Raffelt:1996wa}
\bibitem{Raffelt:1996wa} 
  G.~G.~Raffelt,
  ``Stars as laboratories for fundamental physics : The astrophysics of neutrinos, axions, and other weakly interacting particles,''
  Chicago, USA: Univ. Pr. (1996) 664 p.
  
  \bibitem{iben}
  I.~Iben, Nature {\bf 220}, 5163 (1968).
  
  \bibitem{rood}
  I.~Iben and R.T.~Rood, Nature {\bf 223}, 5209 (1969).
  
  \bibitem{buzzoni}
   A.~Buzzoni \emph{et al.}, Astron.\ Astrophys.\ {\bf 128}, 94 (1983).  
   
   
\bibitem{raffeltry}
 G.~G.~Raffelt,  Astrophys.\ J.\ {\bf 365}, 559 (1990).   
   
%\cite{Friedland:2012hj}
\bibitem{Friedland:2012hj} 
  A.~Friedland, M.~Giannotti and M.~Wise,
  %``Constraining the Axion-Photon Coupling with Massive Stars,''
  Phys.\ Rev.\ Lett.\  {\bf 110}, 061101 (2013)
  [arXiv:1210.1271 [hep-ph]].
  %%CITATION = ARXIV:1210.1271;%%

%\cite{Viaux:2013lha}
\bibitem{Viaux:2013lha} 
  N.~Viaux, M.~Catelan, P.~B.~Stetson, G.~Raffelt, J.~Redondo, A.~A.~R.~Valcarce and A.~Weiss,
  %``Neutrino and axion bounds from the globular cluster M5 (NGC 5904),''
  Phys.\ Rev.\ Lett.\  {\bf 111}, 231301 (2013)
  [arXiv:1311.1669 [astro-ph.SR]].
  %%CITATION = ARXIV:1311.1669;%%

  
  %\cite{Olive:1994fe}
\bibitem{Olive:1994fe} 
  K.~A.~Olive and G.~Steigman,
  %``On the abundance of primordial helium,''
  Astrophys.\ J.\ Suppl.\  {\bf 97}, 49 (1995)
  [astro-ph/9405022].
  %%CITATION = ASTRO-PH/9405022;%%
  
  
  %\cite{Izotov:2013waa}
\bibitem{Izotov:2013waa} 
  Y.~I.~Izotov, G.~Stasinska and N.~G.~Guseva,
  %``Primordial 4He abundance: a determination based on the largest sample of HII regions with a methodology tested on model HII regions,''
  Astron.\ \& Astroph.\ {\bf 558}, A57 (2013)
  [arXiv:1308.2100 [astro-ph.CO]].
  %%CITATION = ARXIV:1308.2100;%%
  
  
  %\cite{Salaris:2004xd}
\bibitem{Salaris:2004xd} 
  M.~Salaris, M.~Riello, S.~Cassisi and G.~Piotto,
  %``The Initial helium abundance of the Galactic GC system,''
  Astron.\ and Astrophys.\  {\bf 420}, 911 (2004)
  [astro-ph/0403600].
  %%CITATION = ASTRO-PH/0403600;%%
  
   %\cite{Gratton:2012ap}
\bibitem{Gratton:2012ap} 
  R.~Gratton, E.~Carretta and A.~Bragaglia,
  %``Multiple populations in GCs. Lessons learned from the Milky Way GCs,''
  Astron.\ and  Astrophys.\ Rev.\ {\bf 20}, 50 (2012)
  [arXiv:1201.6526 [astro-ph.SR]].
  %%CITATION = ARXIV:1201.6526;%%
  
  %\cite{Lee:1994pd}
\bibitem{Lee:1994pd} 
  Y.~-W.~Lee, P.~Demarque and R.~Zinn,
  %``The horizontal-branch stars in GCs. 2: The second parameter phenomenon,''
  Astrophys.\ J.\  {\bf 423}, 248 (1994).
  %%CITATION = ASJOA,423,248;%%
  
%\cite{Straniero:2005hc}
\bibitem{Straniero:2005hc} 
  O.~Straniero, R.~Gallino and S.~Cristallo,
  %``s-Process in low-mass asymptotic giant branch stars,''
  Nucl.\ Phys.\ A {\bf 777}, 311 (2006)
  [astro-ph/0501405].
  %%CITATION = ASTRO-PH/0501405;%% 
  
%\cite{Luciano:2013rya}
\bibitem{Luciano:2013rya} 
  L.~Piersanti, S.~Cristallo and O.~Straniero,
  %``The effects of rotation on the s-process nucleosynthesys in Asymptotic Giant Branch stars,''
  arXiv:1307.2017 [astro-ph.SR].
  %%CITATION = ARXIV:1307.2017;%% 
  
   
%\cite{straniero2014} 
\bibitem{straniero2014} 
  O.~Straniero, S.~Cristallo, and L.~Piersanti, Astrophys.\ J.\ 
  {\bf 785}, 77 (2014)
  [arXiv:1403.0819 [astro-ph.SR]].
  
%\cite{Aver:2013wba}
\bibitem{Aver:2013wba} 
  E.~Aver, K.~A.~Olive, R.~L.~Porter and E.~D.~Skillman,
  %``The primordial helium abundance from updated emissivities,''
  JCAP {\bf 1311}, 017 (2013)
  [arXiv:1309.0047 [astro-ph.CO]].
  %%CITATION = ARXIV:1309.0047;%%

%\cite{Iocco:2008va}
\bibitem{Iocco:2008va} 
  F.~Iocco, G.~Mangano, G.~Miele, O.~Pisanti and P.~D.~Serpico,
  %``Primordial Nucleosynthesis: from precision cosmology to fundamental physics,''
  Phys.\ Rept.\  {\bf 472}, 1 (2009)
  [arXiv:0809.0631 [astro-ph]].
  
  %\cite{Ade:2013zuv}
\bibitem{Ade:2013zuv} 
  P.~A.~R.~Ade {\it et al.}  [Planck Collaboration],
  %``Planck 2013 results. XVI. Cosmological parameters,''
  arXiv:1303.5076 [astro-ph.CO].
  %%CITATION = ARXIV:1303.5076;%%


  
  %\cite{Piersanti:2006zh}
\bibitem{Piersanti:2006zh} 
  L.~Piersanti, O.~Straniero and S.~Cristallo,
  %``A Method to Derive the Absolute Composition of the Sun, the Solar System and the Stars,''
  %Submitted to: Astron.Astrophys.
  Astron.\ and Astrophys.\  {\bf 462}, 1051 (2007)
  [astro-ph/0611229].
   
   %\cite{Serenelli:2010fk}
\bibitem{Serenelli:2010fk} 
  A.~Serenelli and S.~Basu,
  %``Determining the initial helium abundance of the Sun,''
  Astrophys.\ J.\  {\bf 719}, 865 (2010)
  [arXiv:1006.0244 [astro-ph.SR]].
  %%CITATION = ARXIV:1006.0244;%%
  

  
  
  %\cite{Asztalos:2009yp}
\bibitem{Asztalos:2009yp} 
  S.~J.~Asztalos {\it et al.}  [ADMX Collaboration],
  %``A SQUID-based microwave cavity search for dark-matter axions,''
  Phys.\ Rev.\ Lett.\  {\bf 104}, 041301 (2010)
  [arXiv:0910.5914 [astro-ph.CO]].
  %%CITATION = ARXIV:0910.5914;%%
  

\bibitem{Brockway:1996yr}
  J.~W.~Brockway, E.~D.~Carlson and G.~G.~Raffelt,
%  ``SN 1987A gamma-ray limits on the conversion of pseudoscalars,''
  Phys.\ Lett.\ B {\bf 383}, 439 (1996).
%  [astro-ph/ 9605197].

\bibitem{Grifols:1996id}
  J.~A.~Grifols, E.~Mass\'o and R.~Toldr\`a,
%  ``Gamma rays from SN~1987A due to pseudoscalar conversion,''
  Phys.\ Rev.\ Lett.\ {\bf 77}, 2372 (1996).
%  [astro-ph/9606028].
  %%CITATION = ASTRO-PH 9606028;%%
  
% % % % % % % % % %
%On more conventional explanation of Universe transparency to gamma-ray 
% % % % % % % % % %

%\cite{Essey:2009zg}
\bibitem{Essey:2009zg} 
  W.~Essey and A.~Kusenko,
  %``A new interpretation of the gamma-ray observations of active galactic nuclei,''
  Astropart.\ Phys.\  {\bf 33}, 81 (2010)
  [arXiv:0905.1162 [astro-ph.HE]].
  %%CITATION = ARXIV:0905.1162;%%
  %77 citations counted in INSPIRE as of 07 Aug 2014
  
%\cite{Essey:2009ju}
\bibitem{Essey:2009ju} 
  W.~Essey, O.~E.~Kalashev, A.~Kusenko and J.~F.~Beacom,
  %``Secondary photons and neutrinos from cosmic rays produced by distant blazars,''
  Phys.\ Rev.\ Lett.\  {\bf 104}, 141102 (2010)
  [arXiv:0912.3976 [astro-ph.HE]].
  %%CITATION = ARXIV:0912.3976;%%
  %63 citations counted in INSPIRE as of 08 Aug 2014



% % % % % % % % % % % % % % % % % % % % % % % % % % % % % % % % % %
% % % % % % % % % % % % % % % % % % % % % % % % % % % % % % % % % %  

   
   %\cite{Horns:2012fx}
\bibitem{Horns:2012fx} 
  D.~Horns and M.~Meyer,
  %``Indications for a pair-production anomaly from the propagation of VHE gamma-rays,''
  JCAP {\bf 1202}, 033 (2012)
  [arXiv:1201.4711 [astro-ph.CO]].
  %%CITATION = ARXIV:1201.4711;%% 
  
  %\cite{Meyer:2013pny}
\bibitem{Meyer:2013pny} 
  M.~Meyer, D.~Horns and M.~Raue,
  %``First lower limits on the photon-axion-like particle coupling from very high energy gamma-ray observation,''
  Phys.\ Rev.\ D {\bf 87}, 035027 (2013)
  [arXiv:1302.1208 [astro-ph.HE]].
  %%CITATION = ARXIV:1302.1208;%%
  
 

%\cite{Bahre:2013ywa}
\bibitem{Bahre:2013ywa} 
  R.~Bähre, B.~Döbrich, J.~Dreyling-Eschweiler, S.~Ghazaryan, R.~Hodajerdi, D.~Horns, F.~Januschek and E.~-A.~Knabbe {\it et al.},
  %``Any light particle search II ?Technical Design Report,''
  JINST {\bf 8}, T09001 (2013)
  [arXiv:1302.5647 [physics.ins-det]].
  %%CITATION = ARXIV:1302.5647;%%
  %26 citations counted in INSPIRE as of 14 Jun 2014
   
 
%\cite{Irastorza:2011gs} 
\bibitem{Irastorza:2011gs}
 I.~G.~Irastorza {\it et al.},
 %``Towards a new generation axion helioscope,''
 JCAP {\bf 1106}, 013 (2011).
% [arXiv:1103.5334 [hep-ex]].
 %%CITATION = ARXIV:1103.5334;%%


%................Ref. Supplementary material.............................


%\cite{Lemut:2006va}
\bibitem{Lemut:2006va} 
  A.~Lemut {\it et al.}  [LUNA Collaboration],
  %``First measurement of the N-14(p,gamma)O-15 cross section down to 70-KeV,''
  Phys.\ Lett.\ B {\bf 634}, 483 (2006)
  [nucl-ex/0602012].
  %%CITATION = NUCL-EX/0602012;%%
  
  
%\cite{Angulo:1999zz}
\bibitem{Angulo:1999zz} 
  C.~Angulo, M.~Arnould, M.~Rayet, P.~Descouvemont, D.~Baye, C.~Leclercq-Willain, A.~Coc and S.~Barhoumi {\it et al.},
  %``A compilation of charged-particle induced thermonuclear reaction rates,''
  Nucl.\ Phys.\ A {\bf 656}, 3 (1999).
  %%CITATION = NUPHA,A656,3;%%
  
 %\cite{Fynbo:2005vf}
\bibitem{Fynbo:2005vf} 
  H.~O.~U.~Fynbo {\it et al.}  [ISOLDE Collaboration],
  %``Revised rates for the stellar triple alpha process from measurement of C-12 nuclear resonances,''
  Nature {\bf 433}, 136 (2005).
  %%CITATION = NATUA,433,136;%%
  
 %\cite{Schurmann:2012zz}
\bibitem{Schurmann:2012zz} 
  D.~Schurmann, L.~Gialanella, R.~Kunz and F.~Strieder,
  %``The astrophysical S factor of C-12(alpha,gamma)O-16 at stellar energy,''
  Phys.\ Lett.\ B {\bf 711}, 35 (2012).
  %%CITATION = PHLTA,B711,35;%% 
   
 %\cite{Kunz:2001zz}
\bibitem{Kunz:2001zz} 
  R.~Kunz, M.~Jaeger, A.~Mayer, J.~W.~Hammer, G.~Staudt, S.~Harissopulos and T.~Paradellis,
  %``C-12 (alpha, gamma) O-16: The Key Reaction in Stellar Nucleosynthesis,''
  Phys.\ Rev.\ Lett.\  {\bf 86}, 3244 (2001).
  %%CITATION = PRLTA,86,3244;%%  
  
%\cite{Straniero:2003ax}
\bibitem{Straniero:2003ax} 
  O.~Straniero, I.~Dominguez, S.~Cristallo and R.~Gallino,
  %``Low mass agb stellar models for 0.003 <= z <= 0.02: basic formulae for nucleosynthesis calculations,''
  %Submitted to: Publ.Astron.Soc.Austral.
  Publ.\ Astron.\ Soc.\ Austral.\ {\bf 20}, 389 (2003)
  [astro-ph/0310826].
  
 %\cite{Metcalfe:2003ti}
\bibitem{Metcalfe:2003ti} 
  T.~S.~Metcalfe,
  %``White dwarf asteroseismology and the ^12C(alpha,gamma)^16O rate,''
  Astrophys.\ J.\  {\bf 587}, L43 (2003)
  [astro-ph/0303039].
  %%CITATION = ASTRO-PH/0303039;%%
 
 
 %\cite{Esposito:2003wv}
\bibitem{Esposito:2003wv} 
  S.~Esposito, G.~Mangano, G.~Miele, I.~Picardi and O.~Pisanti,
  %``Neutrino energy loss rate in a stellar plasma,''
  Nucl.\ Phys.\ B {\bf 658}, 217 (2003)
  [astro-ph/0301438].
  %%CITATION = ASTRO-PH/0301438;%%
  
  %\cite{Haft:1993jt}
\bibitem{Haft:1993jt} 
  M.~Haft, G.~Raffelt and A.~Weiss,
  %``Standard and nonstandard plasma neutrino emission revisited,''
  Astrophys.\ J.\  {\bf 425}, 222 (1994)
  [Erratum-ibid.\  {\bf 438}, 1017 (1995)]
  [astro-ph/9309014].
  %%CITATION = ASTRO-PH/9309014;%%
   
  %\cite{itoh}
  \bibitem{itoh}
N.~Itoh, A.~Nishikawa, and Y.~Kohyama, Astropys.\ J.\ {\bf 470},
1015 (1996).
%....................................................  







\end{thebibliography}
\end{document}